\documentstyle[prl,aps]{revtex}
\begin{document}
\input{psfig}
\twocolumn[\hsize\textwidth\columnwidth\hsize\csname @twocolumnfalse\endcsname
\draft
\begin{title}
{Tunable adsorption on carbon nanotubes}
\end{title}

\author{O. G\"{u}lseren$^{(1,2)}$, T. Yildirim$^{(1)}$,
and S. Ciraci$^{(3,4)}$}
\address{$^{(1)}$ NIST Center for Neutron Research,
National Institute of Standards and Technology,
Gaithersburg, MD 20899-8562}
\address{$^{(2)}$ Department of Materials Science and Engineering,
University of Pennsylvania, Philadelphia, PA 19104}
\address{$^{(3)}$ Department of Physics, Bilkent University,
Ankara 06533, Turkey}
\address{$^{(4)}$ Department of Physics,
University of Illinois at Chicago,
Chicago, IL 60607-7059}
\date{\today}
\maketitle

\begin{abstract}
We investigated the adsorption of a single atom, hydrogen and aluminum,
on single wall carbon nanotubes from first-principles. The adsorption is
exothermic, and the associated binding energy varies inversely as the
radius of the zigzag tube. We found that the adsorption of a single atom
and related properties can be modified continuously and reversibly by the
external radial deformation. The binding energy on the high curvature site
of the deformed tube increases with increasing radial deformation. The
effects of curvature and radial deformation depend on the chirality of the
tube.
\end{abstract}



]

Novel mechanical, electrical and chemical properties of carbon
nanotubes~\cite{iijima,dressel,bezryadin} have been explored
actively with a motivation of finding a new technological application.
A single wall carbon nanotube (SWNT) is usually described by a rolled
graphene, where the hexagonal 2D lattice is mapped on a cylinder of
radius $R$ with various helicities characterized by a set of two integers
$(n,m)$. A SWNT can display either metallic or insulating electronic
structure depending on the helicity and radius~\cite{dressel}.

Recent studies~\cite{bezryadin,rochef,park00,cetin,tubeprb} showed that
the electronic properties of SWNTs can be modified by radial deformation.
The energy gap of an insulating SWNT can decrease and eventually vanish
at an insulator--metal transition with increasing applied radial strain.
The density of states at the Fermi energy, ${\cal D}(E_F)$, of a metallized
SWNT increases with the increasing radial strain. More interestingly, the
radial deformation necessary to induce metallicity was found to be in the
elastic range. Therefore, all strain induced changes in the electronic and
also in mechanical properties are reversible.

Most noticeably, the radial strain disturbs the uniformity of charge
distribution. This, in turn, may impose changes in the chemical reactivity
and hence in the interaction of tube surface with foreign atoms and
molecules. It is therefore anticipated that not only band gap but also
chemical reactions taking place on the surface of a SWNT can be engineered
through radial deformation. In this paper, we explore this feature by using
the predictive power of the density functional theory and demonstrate that
indeed adsorption of foreign atoms on carbon nanotubes and associated
properties can be modified continuously and reversibly. Furthermore, we
showed that there is a simple scaling of the adsorption energies with the
radius of the SWNT. We believe that the tunable adsorption can have
important implications for metal coverage and selective adsorption
of foreign atoms and molecules on the carbon nanotubes, and can lead to a
wide variety of technological applications, ranging from hydrogen storage
to new materials~\cite{h2store}.

We have investigated the electronic energy structure and charge density
of bare and single atom adsorbed SWNTs with and without radial strain
by performing first-principle density functional calculations~\cite{castep}.
We expressed the wave functions by linear combinations of plane waves up
to an energy cutoff of 500 eV within the tetragonal supercell geometry
(separation between the tubes is taken to be 7.5~\AA{} and $c$ is equal to
the lattice constant of SWNT along its axis). With this energy cutoff
and using ultra soft pseudopotentials for carbon~\cite{usps} the total
energies of nanotube-adatom systems converge within 0.5 meV/atom.
The Monkhorst-Pack~\cite{monpack} special {\bf k}-point scheme is used
with 0.02 \AA{}$^{-1}$ k-point spacing resulting 5 {\bf k}-points along
the tube axis. Results have been obtained within the generalized gradient
approximation~\cite{gga} (GGA) for fully relaxed geometries including all
carbon and adsorbate positions and the lattice constant of the tube along
the $z$-axis. 

\begin{figure}
\centerline{\psfig{figure=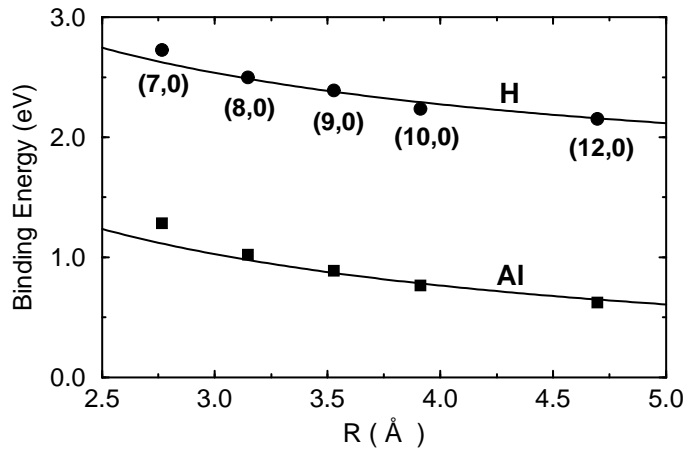,angle=0,width=90mm}}
\vspace*{0.2cm}
\caption{\small Binding energies $E_b$ of single hydrogen and aluminum atom
adsorbed on the zigzag SWNTs versus the radius of the tubes. The solid
line is the fit to $E_b = E_{o} + C/R$ (see text).}
\label{fig:bevsr}
\end{figure}

To investigate the effect of the radial deformation we first consider
adsorption of H and a simple metal, Al, on undeformed SWNTs. In
Fig.~\ref{fig:bevsr}, we present the binding energies of H and Al adsorbed
on the $(n,0)$ zigzag nanotubes calculated for $n=$7,8,9,10,12. H is
adsorbed at the top site, {\it i.e.} directly above the C atoms of the tube,
Al favors the hollow site {\it i.e.} above the center of the hexagon as in
the graphite surface. The binding energy $E_b$,
\begin{equation}
E_b= E_T[SWNT] + E_A - E_T[SWNT+A]
\label{eq:binde}
\end{equation}
is calculated in terms of the total energy of SWNT, $E_T[SWNT]$, total
energy of SWNT with an adsorbed atom A, $E_T[SWNT+A]$, and the energy
of the single, free atom, $E_A$.  Here the bare SWNT and SWNT with an
adatom A are free of external stress, and all the atomic coordinates are 
fully relaxed. Moreover, since $E_{b}$ is calculated by using the same
supercell, the spurious adatom-adatom interaction along the tube axis
is substracted. The positive value of $E_b$ indicates that the adsorption
is exothermic and hence stable. It is found that the binding energy of an
adatom decreases with increasing radius (or decreasing curvature) of the
tube, and eventually saturates at a value corresponding to that on graphene
plane. Hence the variation of binding energies of H and Al with the radius
of the zigzag tube fits to the curve given by the expression,
\begin{equation}
E_{b,A}(R) = E_{o,A} + \frac{C_A}{R}
\end{equation}
where $E_{o,A}$ is the binding energy of the adatom A (H or Al) on the
graphene plane. We calculated $E_{o,H}=$1.49 eV, and $E_{o,Al}=-$0.02 eV.
Interestingly, the fitting parameter $C_A$ is found to be independent of
the adatom (H and Al) and is equal to $\sim$ 3.14~eV\AA. Currently, we are
extending our calculations to see if this relation holds for other
adsorbates as well. The binding energies calculated for $(n,0)$ SWNTs with
$n<8$ deviate from the above simple scaling perhaps due to the fact that
the singlet $\pi^{*}$-band, which is normally in the conduction band,
falls into the band gap as a result of increased $\sigma^{*}-\pi^{*}$
mixing at high curvature~\cite{tubeprb,blase}. Note that, while the band gap
shows significant change with $n$, (for example, upon going from $(8,0)$ to
$(9,0)$ $E_{g}$ changes from 0.65 eV to 0.09 eV) the binding energies vary
smoothly with $R^{-1}$. Increasing $E_b$ with decreasing $R$ (or with
decreasing $n$) shows that for small $R$ the character of the surface
deviates from that of the graphene. This finding also suggests that by
creating regions of different curvature on a single SWNT by radial
deformation one can attain different values of binding energies.

The radial deformation that we consider in this study is generated by
applying uniaxial compressive stress on a narrow strip on the surface
of the SWNT~\cite{tubeprb}. In practice such a deformation can be realized
by pressing the tube between two rigid flat surfaces. The radius
is decreased in the $y$-direction, while it is elongated along the
$x$-direction. As a result, the circular cross section is distorted
to an elliptical one with major and minor axis, $a$ and $b$ respectively.
The elliptical radial deformation can be described by the magnitude of
the applied strain and it is defined as $\epsilon_{yy} = (R - b)/R$
where $R$ is the radius of the undeformed nanotube (with zero strain).
For different values of $\epsilon_{yy}$ we carried out full structural
optimization under the constraint that the minor axis is kept fixed
at a preset value by freezing the carbon atoms at both ends of the minor
axis~\cite{tubeprb}. Then, under this constraint, the coordinates of the
adatom and remaining carbon atoms and the lattice constant of the tube,
$c$ are optimized until there remain only forces opposite to the applied
strain on the fixed atoms, but all other force components are less than
0.01 eV/\AA.

\begin{figure}
\centerline{\psfig{figure=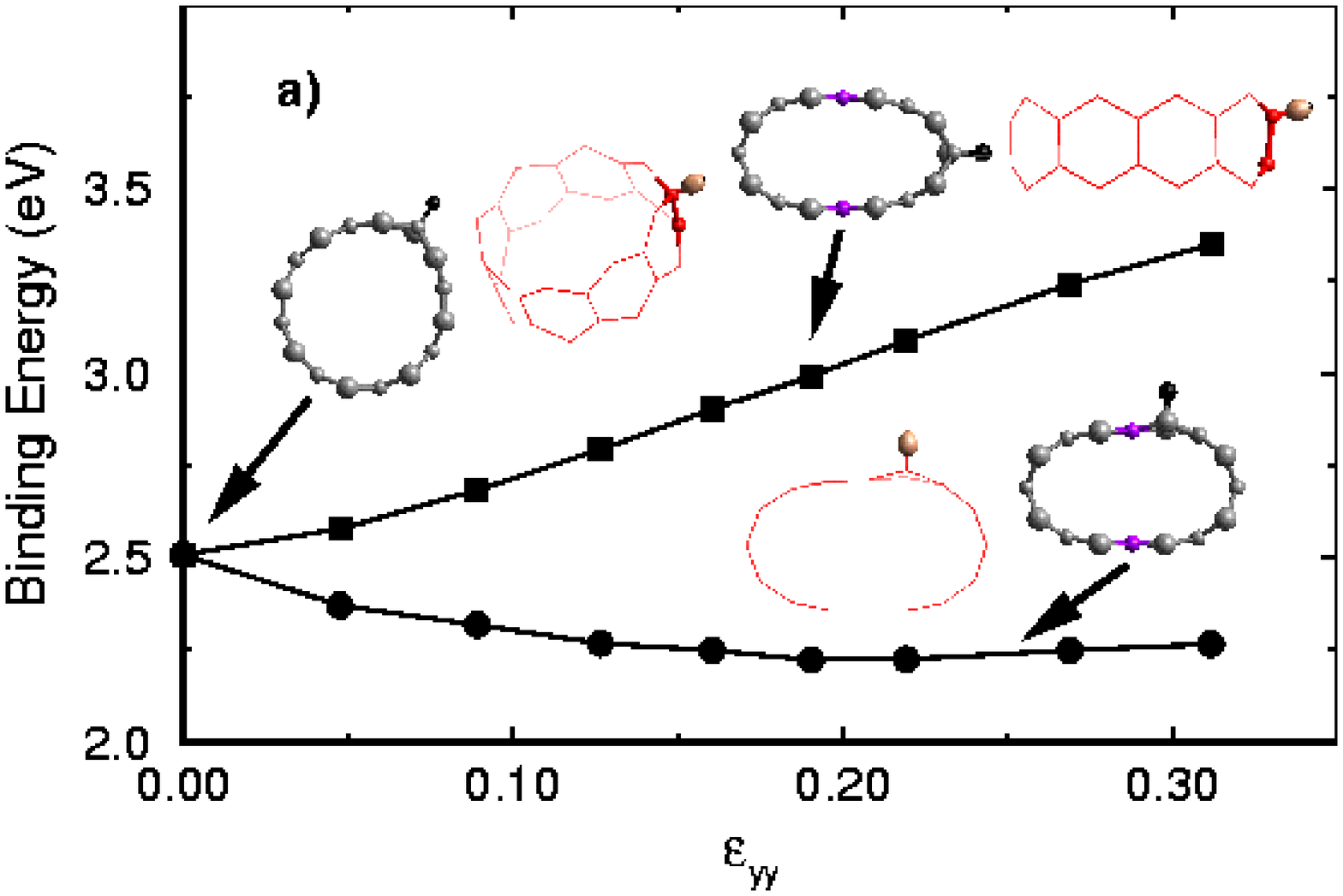,angle=0,width=90mm}}
\vspace*{0.2cm}
\centerline{\psfig{figure=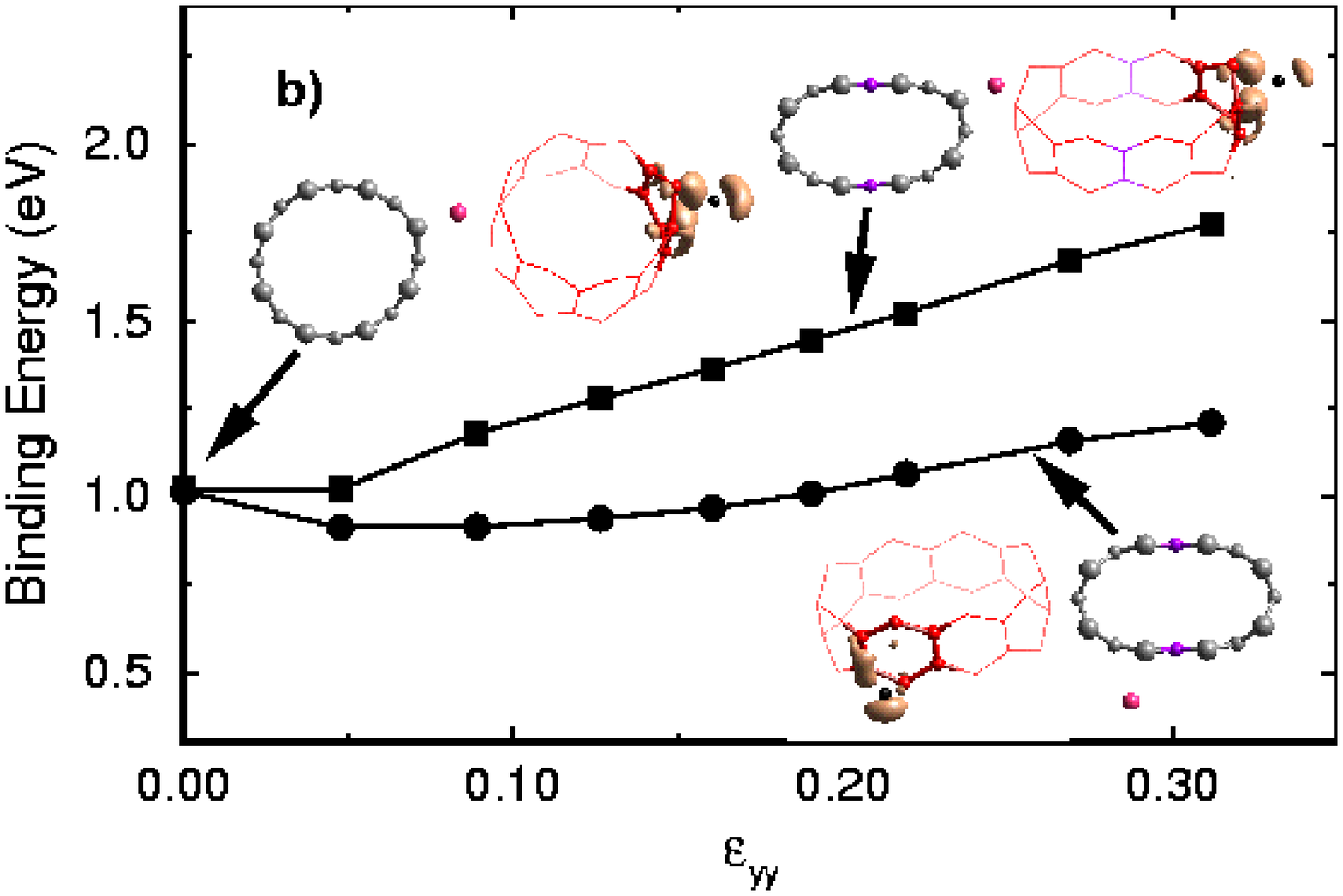,angle=0,width=90mm}}
\caption{\small (a) Variation of the binding energy $E_b$ of a single hydrogen
atom adsorbed on a $(8,0)$ zigzag SWNT as a function of the elliptic radial
deformation, $\epsilon_{yy}$.
The upper curve corresponds to H adsorbed on the high curvature region
near the end of the major axis $a$ ({\it sharp site}). The lower curve is
for the adsorption on the low curvature region at the end of the minor axis
({\it flat site}) (b) Same as (a) for a single adsorbed Al atom.
Insets: Ball and stick models and isosurface plot of difference charge
densities, $\Delta \rho$ for each case. (See text).}
\label{fig:strbe}
\end{figure}

Figure~\ref{fig:strbe}a shows the variation of the binding energy $E_b$
of a single hydrogen atom adsorbed on the $(8,0)$ surface with the applied
radial strain for two cases: (i) The binding energy of a single H atom
adsorbed on the high curvature side of the surface ({\it i.e.} specified
as the {\it sharp-site} near one of the ends of the major axis, $x=a$,
$y=0$) traces the upper curve in Fig. 2a.  (ii) The lower curve is for the
adsorption on the low curvature side near one of the ends of the minor axis
({\it i.e.} $x=0$, $y=b$ specified as the {\it flat-site}). Here, the
binding energy is defined as in Eq(1), except that $E_T[SWNT]$ and
$E_T[SWNT+A]$ are calculated for radially deformed SWNTs. The binding
energy of H adsorbed on the high curvature (sharp) site is increased by
0.85 eV for $\epsilon_{yy}$=0.3. On the other hand, $E_b$ for the adsorption
on the low curvature (flat) site first decreases with increasing
$\epsilon_{yy}$, and then saturates at an energy 0.25 eV less than that
corresponds to $\epsilon_{yy}=$0.0. The difference of binding energies of
the sharp and flat sites, $\Delta E_b$, is substantial and is equal to
$\sim$ 1.1 eV. This value is 44 \% of the binding energy of H on the
undeformed SWNT. As a result of H adsorption, the $sp^2$ character of the
bonding of the tube has changed locally and became more like to $sp^3$. The
lengths of the C--C bonds at the close proximity of H have increased
slightly.

The binding energy of Al shown in Fig.~\ref{fig:strbe}b exhibits a behavior
similar to that of H, despite H and Al favor different sites on the $(8,0)$
tube: $E_b$ at the sharp site of the deformed SWNT increases with increasing
$\epsilon_{yy}$. For example, $E_b$ increases by $\sim$ 0.80 eV for
$\epsilon_{yy}=0.3$ which is 80\% of the binding energy on the undeformed
tube. For Al absorbed on the flat site, $E_b$  first decreases with
increasing $\epsilon_{yy}$, then gradually increases. Adsorption of Al
induces local changes in the atomic and electronic structure. For example,
the surface of the tube where Al is adsorbed expands.

The variation of $E_b$ with the radial deformation is consistent with the
results illustrated in Fig.~\ref{fig:bevsr}. In general, the higher is the
curvature under deformation, the higher the binding energy. The effect of
the elastic deformation is further investigated by analyzing the charge
density and electronic structure. The Mulliken analysis estimates that
$\sim$0.37 electrons is transfered from H to carbon~\cite{tubech}.
The C--H bond is directional and is covalent with a partial ionic component.
The charge difference, $\Delta \rho=\rho[SWNT+H]-\rho[SWNT]-\rho[H]$, which
is calculated in terms of the undeformed (or deformed) SWNT with H (adsorbed
at different sites), undeformed (deformed) clean SWNT, and single H, are
shown by inset in Fig.~\ref{fig:strbe}a. $\Delta \rho$ indicates no
significant change with radial deformation. On the other hand, the charge
transfer upon the adsorption of Al is different from H. Since Al is adsorbed
on the center of a hexagon, the bond between Al and SWNT is distributed to
nearest C atoms with charge accumulation between Al and those C atoms. The
bond charge slightly increases at the sharp site as shown by inset in
Fig.~\ref{fig:strbe}b. The charge transfer from Al atom is estimated to be
0.71 electrons.

\begin{figure}
\centerline{\psfig{figure=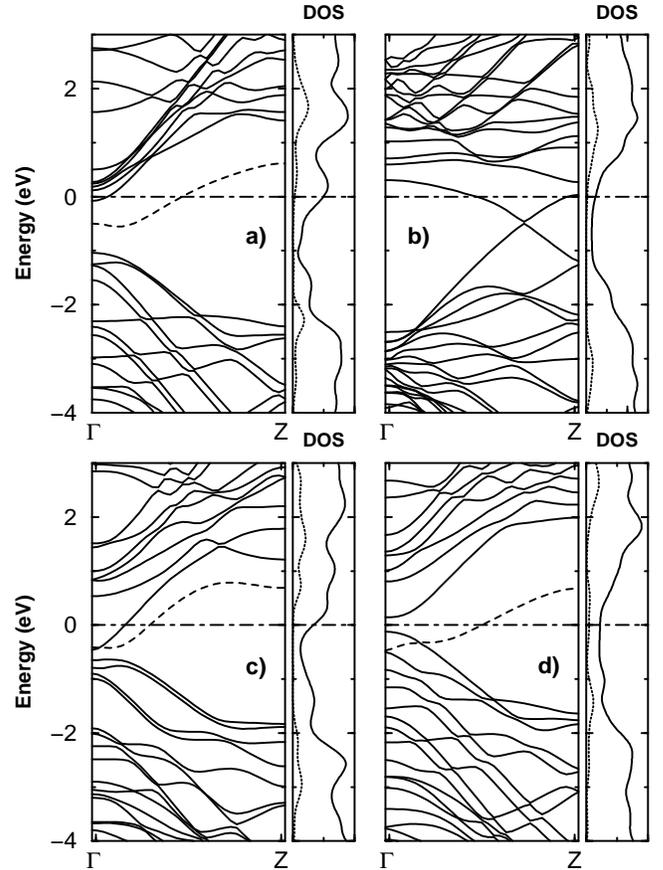,angle=0,width=90mm}}
\vspace*{0.2cm}
\caption{\small Electronic band structure, total and partial density of states
(with dotted lines) of Al atom adsorbed on a SWNT. The zero of energy is
taken at the Fermi level shown by dash-dotted lines. The band due to Al
in the band gap is shown by dashed line. Al adsorbed: (a) on the undeformed
$(8,0)$ tube, (b) on the undeformed $(6,6)$ tube, (c) on the flat site of
the $(8,0)$ tube, (d) on the sharp site of the $(8,0)$ tube.
$\epsilon_{yy}=0.31$ for (c) and (d).}
\label{fig:bands}
\end{figure}

Explanation of this remarkable and significant change of the binding
energy with radial deformation is sought in the electronic energy structure
and the total and partial density of states. The adsorption of H gives rise
to a new state which falls in the gap at $\Gamma$ and coincides with $E_F$.
This state partially overlaps with the conduction band when H is adsorbed
at the flat site, whereas it occurs near the valence band edge at the sharp
site. A similar situation occurs with the adsorption of Al as illustrated
in Fig.~\ref{fig:bands}. The band gap is wide open, and a resonant state
$\sim$ 2.5 eV below $E_{F}$ and a $p$-derived  adsorption state which sets
$E_{F}$ falls at the center of the gap for the adsorption of Al on the
undistorted SWNT~\cite{disper}. In the case of adsorption on the flat site
the latter state overlaps the conduction band (Fig.~\ref{fig:bands}c), but
it dips into the valence band at the sharp site (Fig.~\ref{fig:bands}d).
Accordingly, the sharp-site adsorption involves the valence band states of
SWNT, while the conduction band states are involved in the flat site
adsorption.

\begin{figure}
\centerline{\psfig{figure=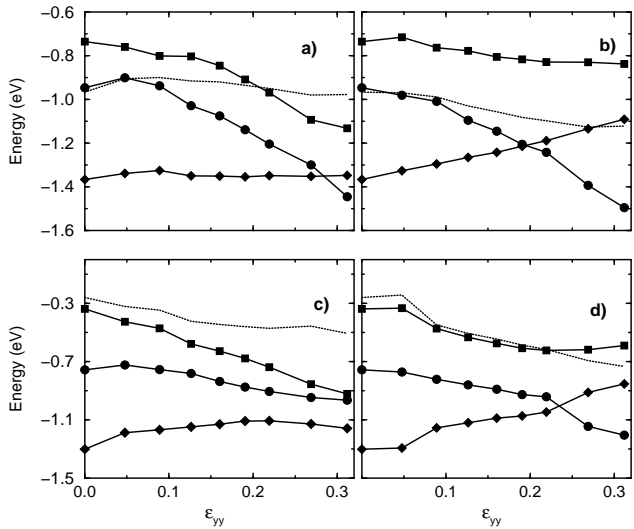,angle=-90,width=90mm}}
\vspace*{0.2cm}
\caption{\small The shifts of band edges at $\Gamma$-point with radial
deformation on a $(8,0)$ tube. H adsorbed: (a) on the flat site, (b) on the
sharp site; Al adsorbed: (c) on the flat site, (d) on the sharp
site. The valence band edge, the conduction band edge, the adsorption state
and the Fermi energy are shown by diamonds, squares, circles and dots,
respectively.}
\label{fig:edge}
\end{figure}

The above argument is in accord with the binding energies calculated for
the adsorption of Al on the undeformed and deformed $(6,6)$ armchair SWNT.
Note that the $(6,6)$ as well as all $(n,n)$ SWNTs are metallic
with finite ${\cal D}(E_F)$ owing to the $\pi-$ and $\pi^{*}-$states
crossing and setting $E_F$. Our calculations show that ${\cal D}(E_F)$
of the $(6,6)$ decreases slightly with increasing radial deformation,
but it remains essentially finite. We found $E_b=$0.91 eV for the
adsorption of Al on the $(6,6)$ SWNT. Upon the deformation of the tube
with $\epsilon_{yy}=$0.22, $E_b$ corresponding to sharp-, and flat-site
adsorption has changed to 0.95 and 0.85 eV, respectively. Therefore, the
difference between the binding energies of  the sharp and flat site
adsorption is only 0.1 eV, and hence is negligible as compared to the
situation above discussed for the $(8,0)$ SWNT. In Fig.~\ref{fig:bands}b
the state associated with the adsortion of Al on the undeformed $(6,6)$
tube occurs above the Fermi level and donates its electron to the metallic
$\pi-$ and $\pi^{*}-$bands. It appears that $E_b$ is almost pinned by the
above mechanism.

Radial deformation induced changes in the energies of valance band edge
(VBE) and conduction band edge (CBE), as well as the adsorbtion state (AE)
at $\Gamma$--point illustrated in Fig.~\ref{fig:edge} corraborate the above
arguments. In the case of adsorption on the flat site, VBE is almost constant
for both H and Al, while CBE and AE both shift downwards. As a result of this,
more conduction band states contribute to the bond energy. By contrast,
for adsorption on the sharp site, while AE behaves very similar to the
previous case, CBE is now approximately constant, but VBE shifts upwards.
In this situation, the contribution to the bond energy from conduction
band states is decreased.

In conclusion, we showed that the chemical reactivity of a zigzag SWNT
can be modified reversibly and variably by the radial deformation.
The effect of deformation is significantly different for the zigzag and
armchair SWNTs. It is remarkable that Al, which is not bound to the graphite
surface, can be adsorbed at a high curvature site of a zigzag SWNT under
radial deformation with a binding energy of 1.8 eV. This novel property may
have important implications for various chemical and electronic applications,
such as selective absorption and desorption of molecules and atoms,
fragmentation and chemical sensors, magnetic tubes {\it etc}.

{\bf Acknowledgements}
This work was partially supported by the NSF under Grant No. INT97-31014
and T\"{U}B\'{I}TAK under Grant No. TBAG-1668(197 T 116).
SC thanks Prof. S. S\"{u}zer for stimulating discussions.


\begin{references}

\bibitem{iijima} 
S. Iijima,
{\sl Nature} {\bf 354}, 56 (1991).

\bibitem{dressel} 
M. S. Dresselhaus, G. Dresselhaus and P. C. Eklund,
{\sl Science of Fullerenes and Carbon Nanotubes};
Academic Press, San Diego (1996);
R. Saito, G. Dresselhaus, and M. S. Dresselhaus, 
{\sl Physical Properties of Carbon Nanotubes};
Imperial College Press, London (1998);
N. Hamada, S. Sawada and A. Oshiyama,
{\sl Phys. Rev. Lett.} {\bf 68}, 1579 (1992).

\bibitem{bezryadin}
A. Bezryadin, A. R. M. Verschueren, S. J. Tans, and C. Dekker,
{\sl Phys. Rev. Lett.} {\bf 80}, 4036 (1998).

\bibitem{rochef} 
A. Rochefort, P. Avouris, F. Lesage, D. R. Salahub,
{\sl Chem. Phys. Lett.} {\bf 297}, 45 (1998).

\bibitem{park00} 
C. J. Park, Y. H. Kim, K. J. Chang, 
{\sl Phys. Rev. B} {\bf 60}, 10656 (1999).

\bibitem{cetin} 
\c{C}. K{\i}l{\i}\c{c}, S. Ciraci, O. G\"{u}lseren, and T. Yildirim,
{\sl Phys. Rev. B} {\bf 62}, 16345 (2000).

\bibitem{tubeprb}
O. G\"{u}lseren, T. Yildirim, S. Ciraci and \c{C}. K{\i}l{\i}\c{c},
submitted to {\sl Phys. Rev. B} (2001).

\bibitem{h2store} 
A.C. Dillon {\it et. al.},
{\sl Nature} {\bf 386}, 377 (1997);
C. Liu, {\it et. al.},
{\sl Science} {\bf 286}, 1127 (1999);
Q. Wang, {\it et. al.},
{\sl Phys. Rev. Lett.} {\bf 82}, 956 (1999);
M.C. Gordillo, J. Boronat and J. Casulleras,
{\sl Phys. Rev. Lett.} {\bf 85} 2348 (2000);
M.M. Calbi, F. Toigo and M.W. Cole,
{\sl Phys. Rev. Lett.} {\bf 86} 5062 (2001).

\bibitem{castep} 
M.C. Payne {\it et. al.}, 
{\sl Rev. Mod. Phys.} {\bf 64}, 1045 (1992).

\bibitem{usps} 
D. Vanderbilt,
{\sl Phys. Rev. B} {\bf 41}, 7892 (1990).

\bibitem{monpack} 
H.J. Monkhorst and J.D. Pack,
{\sl Phys. Rev. B} {\bf 13}, 5188 (1976).

\bibitem{gga} 
J. P. Perdew and Y. Wang,
{\sl Phys. Rev. B} {\bf 46}, 6671 (1992).

\bibitem{blase} 
X. Blase, L.X. Benedict, E.L. Shirley and S.G. Louie,
{\sl Phys. Rev. Lett.} {\bf 72}, 1878 (1994).

\bibitem{tubech} 
T. Yildirim, O. G\"{u}lseren, and S. Ciraci,
{\sl Phys. Rev. B.} {\bf 64}, 075404 (2001).

\bibitem{disper}
Note that the band associated the adsorption state in the band gap
displays a small dispersion for Al but practically no dispersion for H.
This is the artifact of the supercell geometry used in the present study.

\end{references}
\end{document}